\documentclass[pra]{revtex4}

\usepackage{amssymb}
\usepackage{graphicx}
\usepackage{amsmath}
\usepackage{amsthm}
\usepackage[ansinew]{inputenc} 
\usepackage{bbm}
\usepackage{epsfig}

\newcommand{\beq}{\begin{eqnarray}}
\newcommand{\eeq}{\end{eqnarray}}
\newcommand{\bra}[1]{\ensuremath{\langle #1 |}}
\newcommand{\ket}[1]{\ensuremath{| #1 \rangle}}

\begin{document}

\begin{flushright}
UWThPh-2011-10
\end{flushright}

\title{A Computable Criterion for Partial Entanglement in Continuous Variable Quantum Systems}

\author{Andreas Gabriel}
\email{Andreas.Gabriel@univie.ac.at}
\affiliation{University of Vienna, Faculty of Physics, Boltzmanngasse 5, 1090 Vienna, Austria}

\author{Marcus Huber}
\email{Marcus.Huber@univie.ac.at}
\affiliation{University of Vienna, Faculty of Physics, Boltzmanngasse 5, 1090 Vienna, Austria}

\author{Sasa Radic}
\email{Sasa.Radic@univie.ac.at}
\affiliation{University of Vienna, Faculty of Physics, Boltzmanngasse 5, 1090 Vienna, Austria}

\author{Beatrix C. Hiesmayr}
\email{Beatrix.Hiesmayr@univie.ac.at}
\affiliation{Research Center for Quantum Information, Institute of Physics,
Slovak Academy of Sciences, Dúbravská cesta 9, 84511 Bratislava, Slovakia and\\ University of Vienna, Faculty of Physics, Boltzmanngasse 5, 1090 Vienna, Austria.}

\begin{abstract}
A general and computable criterion for $k$-(in)separability in continuous multipartite quantum systems is presented. The criterion can be experimentally implemented with a finite and comparatively low number of local observables.  We discuss in detail how the detection quality can be optimised.\end{abstract}

\maketitle

\section{Introduction}
Quantum entanglement has been studied quite intensely over the last few decades, resulting in a rather wide understanding of simple entangled systems (i.e. bipartite systems of low dimensions, especially two qubit systems; for an overview, see e.g. Ref.~\cite{horodeckiqe}). However, still many puzzling features and open questions are revealed in more general systems, such as multipartite systems or systems of high (or, in particular, infinite) dimensions.\\
In multipartite entangled systems (which are of grave importance to technological applications of quantum information theory, such as quantum secret sharing \cite{schauer} or quantum computation \cite{computer}), complications arise (among others) due to the multiple different forms in which a multipartite state can be entangled (see e.g. \cite{huberbruss, relclasses}). In particular, while a bipartite state is either entangled or separable, a multipartite state can be partially entangled (as investigated e.g. in Refs~\cite{k-sep, k-sep2}), as opposed to genuinely multipartite entangled (see e.g. \cite{gme1, gme2, gme3, gme4, gme5}).\\
In infinite-dimensional (continuous) systems, problems arise because certain notions of finite-dimen\-sional systems are no longer met (see e.g. \cite{continuoussummary, continuousbe}). Nevertheless, many concepts have been generalised from the finite-dimensional case to the continuous one during the last decade, most noteworthy the PPT-criterion \cite{ppt} and the scheme of quantum teleportation \cite{teleport} (which has also already been experimentally verified \cite{teleportexp}).\\
Rather seldomly are systems considered which contain both these sets of difficulties (e.g. \cite{giedkemulti, hyllus}, although multipartite continuous quantum systems do hold the possibility for certain applications, such as certain kinds of teleportation networks \cite{multipartcont}.\\
In this letter, we use a general framework (which was introduced in Refs.~\cite{hmgh,ghh} for finite-dimensional systems) to formulate a criterion for partial separability ($k$-separability) of arbitrary states of a continuous variable multipartite system, thus proving implicitly, that the framework also works perfectly well in infinite-dimensional systems.\\
The article is organised as follows. In section \ref{sec_def}, the basic definitions and terminology will be reviewed, such that in section \ref{sec_crit} we can present our criterion for continuous variable $k$-separability, which is the main result of this letter. In section \ref{sec_opt}, a guideline to the application of the criterion will be given (such that its detection power can be optimally used). The criterion and its application are then demonstrated in two exemplary cases in section \ref{sec_examples}. Finally, in section \ref{sec_exp} we show how the criterion can also be implemented experimentally.\\

\section{Basic Definitions\label{sec_def}}
In order to formulate our result, we firstly need to define the concept of $k$-separability for continuous variable systems. A general $n$-partite pure quantum state can be written in the form
\beq \ket{\Psi} = \int_{-\infty}^{\infty} d^{n}x \ \Psi(x_1,x_2,\cdots,x_n) \ \ket{x_1}\otimes\ket{x_2}\otimes\cdots\otimes\ket{x_n}\;. \eeq
It is called $k$-separable ($k \leq n$) iff its distribution function factorises into $k$ factors, i.e.
\beq \Psi(x_1,x_2,\cdots,x_n) = \Psi_1(\{x_{i_1}\})\cdot\Psi_2(\{x_{i_2}\})\cdot \dots\cdot \Psi_k(\{x_{i_k}\}) \eeq
where $\{x_{i_j}\}$ denotes a set of coordinates, i.e. corresponds to one or several particles. That is, a pure state is called $k$-separable iff there is a $k$-partition $(\{x_{i_1}\}|\{x_{i_2}\}|\cdots|\{x_{i_k}\})$ with respect to which it is separable. If $k=n$ then the state is called fully separable, i.e. there is no entanglement in the multipartite system. If $k=2$ the state is called biseparable, if it is not biseparable, it is called genuinely multipartite entangled. Genuine multipartite entanglement is found to be a key ingredient for many quantum algorithms, see e.g. Ref.~\cite{Bruss}. In finite-dimensional systems, there exist different inequivalent classes of genuine multipartite entanglement, e.g. the $GHZ$-class, the $W$-class or the Dicke-class. Such substructures are only known for very special and rather simple systems (see e.g. \cite{classes1,classes2,huberbruss}) and it is not known how this generalises for continuous variable systems.\\
\\
\noindent\textbf{Example:} A tripartite pure state can be either fully separable, i.e. $k=n$ with the $3$-partition $(x_1|x_2|x_3)$, or 2--separable (biseparable, i.e. k=2) with one of the three possible partitions $(x_1 x_2|x_3)$, $(x_1 x_3|x_2)$ or $(x_1|x_2 x_3)$, or genuinely multipartite entangled ($k=1$) with the partition $(x_1 x_2 x_3)$.\\
\\
For mixed states, we can extend this definition in a straightforward way. A mixed $n$-partite quantum state has the general form
\beq \rho = \int_{-\infty}^{\infty} d^{n}x \ d^{n}x' \ \rho({x_1,x_1',x_2,x_2',\dots,x_n,x_n'})\; \ket{x_1}\bra{x_1'}\otimes\ket{x_2}\bra{x_2'}\otimes\cdots\ket{x_n}\bra{x_n'}\eeq
It is called $k$-separable, iff it has a decomposition into $k$-separable pure states, i.e. iff it can be written as a convex combination of pure $k$-separable states:
\beq \rho = \int d\alpha\; p_\alpha\; \ket{\Psi_\alpha}\bra{\Psi_\alpha} \eeq
where $p_\alpha$ is a probability distribution (i.e. $p_\alpha \geq 0$ and $\int_{-\infty}^{\infty} d\alpha \ p_\alpha = 1$) and $\ket{\Psi_\alpha}$ is $k$-separable for all $\alpha$.\\
Note that a $k$-separable mixed state may not be separable with respect to any partition, since the pure states in its decomposition may be separable with respect to different $k$-partitions (i.e. $\ket{\Psi_\alpha}$ may split into different partitions for different $\alpha$). The concept of $k$-separability is of high impoertance to quantum information theory, since many of its applications rely on specific kinds of states (in particular on genuinely multipartite entangled states).\\

\section{Criterion for k-Separability\label{sec_crit}}
In Refs.~\cite{hmgh,ghh}, a framework for constructing very general separability criteria for finite Hilbert spaces was introduced. We now extend this framework such that we can apply it to continuous quantum systems. The main result of this paper is an inequality, given in the following theorem, which is satisfied for all $k$--separable states, such that any violation implies that the state under investigation is not $k$--separable.\\
\\
\textbf{Theorem:} All $k$-separable states $\rho$ satisfy the inequality
\begin{equation} \tag{$\ast$} \sqrt{\bra{\Phi}\rho^{\otimes 2} P_{total}\ket{\Phi}} - \sum_{\{\alpha\}}\left(\prod_{i=1}^k \bra{\Phi}P_{\alpha_i}^\dagger \rho^{\otimes 2} P_{\alpha_i} \ket{\Phi}\right)^{\frac{1}{2k}} \leq 0 \label{ineq} \end{equation}
where $\ket{\Phi}=\ket{\phi_1}\otimes\ket{\phi_2}$ is an arbitrary fully separable state, $P_{\alpha_i}$ is a permutation operator permuting the $\alpha_i$-th elements of $\ket{\phi_1}$ and $\ket{\phi_2}$, $P_{total}$ wholly permutes the $\ket{\phi_i}$ and the sum runs over all $k$-partitions $\{\alpha\}$.\\
\\
\textbf{Proof:} Firstly, observe that the left hand side of the inequality (\ref{ineq}) is a convex quantity. Thus, its validity for mixed states follows from its validity for pure states. To prove the latter, let us assume w.l.o.g. that a given pure state $\rho=\ket{\Psi}\bra{\Psi}$ is separable with respect to the partition $\alpha'$. Therefore, the permutation operators corresponding to this partition do not change the two copy state
\beq P_{\alpha'_i}\ket{\Psi}^{\otimes 2} = P^\dagger_{\alpha'_i}\ket{\Psi}^{\otimes 2} = \ket{\Psi}^{\otimes 2} \ \forall \ i \;.\eeq
Furthermore, note that the total permutation acts as
\beq P_{total}\ket{\phi_1}\otimes\ket{\phi_2} = \ket{\phi_2}\otimes\ket{\phi_1} \eeq
then the inequality reads
\beq \left|\bra{\phi_1}\rho\ket{\phi_2}\right| - \sqrt{\bra{\phi_1}\rho\ket{\phi_1}\bra{\phi_2}\rho\ket{\phi_2}} - \sum_{\{\alpha\neq\alpha'\}} \left(\prod_{i=1}^k \bra{\Phi}P_{\alpha_i}^\dagger \rho^{\otimes 2} P_{\alpha_i} \ket{\Phi}\right)^{\frac{1}{2k}} \leq 0 \;. \eeq
It follows from the Cauchy-Schwarz inequality that the first two terms cancel (because the pure state $\rho$ is a projector). Since the remaining sum has strictly nonnegative terms with an overall negative sign, the whole inequality has to be satisfied for $\rho$ which are separable with respect to the partition $\alpha'$. \qed \\

\section{Optimising the Detection Quality\label{sec_opt}}
Evidently, the detection quality of inequality (\ref{ineq}) strongly depends on the choice of the separable state $\ket{\Phi}$. Unlike in the case of discrete systems, numerical optimisation is quite out of the question, as the number of optimisation parameters would be infinite. However, there is a quite intuitive way of choosing $\ket{\Phi}$ effectively. Given the state $\rho$ in question, $\ket{\Phi}$ should satisfy the following conditions:
\begin{enumerate}
	\item[{\cal C}1:] $\ket{\Phi}$ has to be fully separable, i.e. $\ket{\Phi} = \bigotimes_{i=1}^{n} \ket{\phi_{1i}} \otimes \bigotimes_{i=1}^{n} \ket{\phi_{2i}}$. \label{item_fs}
	\item[{\cal C}2:] $\ket{\phi_1}$ and $\ket{\phi_2}$ should be orthogonal in each subsystem, i.e. $\bra{\phi_{1i}}\phi_{2i}\rangle = 0 \ \forall \ i$. \label{item_ortho}
	\item[{\cal C}3:] Each $\ket{\phi_{ji}}$ should be sharp, i.e. $\ket{\phi_{ij}} = \int_{-\infty}^{\infty}dx \ \delta(x-x_{ij}) \ket{x} = \ket{x_{ij}}$ for some $x_{ij}$. \label{item_sharp}
	\item[{\cal C}4:] $\ket{\phi_1}$ and $\ket{\phi_2}$ should be chosen such that $|\bra{\phi_1}\rho\ket{\phi_2}|$ is maximal. \label{item_offdiag}
\end{enumerate}
Let us illustrate the background of these conditions.\\
Condition ${\cal C}1$ is necessary for the criterion (ineq. (\ref{ineq})) to remain valid for all $k$-separable states and thus is rather a technical requirement.\\
Condition ${\cal C}2$ guarantees that the permutation operators used in the criterion (\ref{ineq}) have maximal impact, such that a maximal violation of the inequality can be achieved. Each pair $\ket{\phi_{ji}} \ (j=1,2)$ is responsible for detecting entanglement in the $i$-th subsystem. Thus, non-orthonormality of this pair of vectors means non-optimal detection of entanglement in this subsystem.\\
Condition ${\cal C}3$ stems from the fact, that any average is always lower than (or equal to) its highest contribution. Any distribution containing more than one element leads to an averaging in the scalar products in (\ref{ineq}), which can never increase violation of the inequality, but will in general decrease it.\\
Condition ${\cal C}4$ incorporates the dependance of $\ket{\Phi}$ on the investigated state and thus most of the subtleties involved. The first term in ineq. (\ref{ineq}) (and the only positive one) is the absolute value of the scalar product $\bra{\phi_1}\rho\ket{\phi_2}$. Evidently, for the inequality to be violated, this product needs to be as big as possible. This corresponds to chosing $\ket{\phi_1}$ and $\ket{\phi_2}$ as two of the main contributing vectors of the investigated state.\\
Although this procedure might not be unique (i.e. might not unambiguously lead to a definite choice of $\ket{\Phi}$), it reveals the optimal structure of $\ket{\Phi}$ and thus drastically reduces the number of optimisation parameters to a finite set, which can be optimised numerically (e.g. by means of the method introduced in \cite{spengler}) or even analytically. This will be demonstrated in the next section.\\

\section{Examples\label{sec_examples}}
Consider the family of tripartite states (i.e. n=3)
\beq \nonumber \label{statedef} \ket{\omega} = \frac{1}{\sqrt{N}} && \int_{-\infty}^{\infty} d^3x \ e^{-\frac{x_1^2}{2\sigma}} e^{-\frac{(x_1-x_2)^2+(x_1-x_3)^2}{2 \epsilon}} \\ && (\ket{x_1-\Delta}\otimes\ket{x_2}\otimes\ket{x_3+\Delta}+\ket{x_1}\otimes\ket{x_2+\Delta}\otimes\ket{x_3-\Delta}+\ket{x_1+\Delta}\otimes\ket{x_2-\Delta}\otimes\ket{x_3}+ \\ \nonumber &&
\ket{x_1+\Delta}\otimes\ket{x_2}\otimes\ket{x_3-\Delta}+\ket{x_1}\otimes\ket{x_2-\Delta}\otimes\ket{x_3+\Delta}+\ket{x_1-\Delta}\otimes\ket{x_2+\Delta}\otimes\ket{x_3}) \eeq
where 
\beq
N = 2 \pi^{\frac{3}{2}} \epsilon \sqrt{\sigma}
 (e^{-\frac{2 \Delta^2}{\epsilon}}+2 e^{-\frac{\Delta^2}{2\epsilon}} + 2 e^{-\frac{\Delta^2 (\epsilon + 5 \sigma)}{\epsilon \sigma}} + 4 e^{-\frac{\Delta^2 (\epsilon + 5 \sigma)}{4 \epsilon \sigma}}+ 2 e^{-\frac{\Delta^2 (\epsilon + 9 \sigma)}{4 \epsilon \sigma}}+ 2 e^{-\frac{\Delta^2 (\epsilon + 18 \sigma)}{4 \epsilon \sigma}} + 2\sqrt{2} e^{\frac{\sigma-2\Delta(\Delta \epsilon + 6 \sigma)}{8 \epsilon \sigma}} )
\eeq
is a normalisation constant and all parameters $\sigma, \epsilon$ and $\Delta$ are larger than or equal to zero.  Note that
\beq \lim_{\epsilon \rightarrow 0} (2 \pi \epsilon)^{-\frac{1}{2}} e^{-\frac{x^2}{2\epsilon}} = \delta(x) \label{deltafkt} \eeq
is the Dirac delta function, such that in this limit, for $\Delta = 0$ the state $\ket{\omega}$ is a generalisation of the $GHZ$-state to continuous systems:
\beq \lim_{\epsilon \rightarrow 0} \ket{\omega}_{\Delta = 0} = \frac{1}{\sqrt{N}} \int_{-\infty}^{\infty} dx \ e^{-\frac{x^2}{2 \sigma}} \ket{x}\otimes\ket{x}\otimes\ket{x} \eeq
while for $\Delta > 0$ it is a generalisation to the W-state, since it is genuinely multipartite entangled and has entangled reduced density matrices.\\
In the following we are interested in the introduced tripartite states mixed with Gaussian distributed noise
\beq \label{rhomix} \rho_{mix} = (2 \pi \delta)^{-\frac{3}{2}}\int_{-\infty}^{\infty} d^3x \ e^{-\frac{x_1^2+x_2^2+x_3^2}{2 \delta}} \ \ket{x_1}\bra{x_1}\otimes\ket{x_2}\bra{x_2}\otimes\ket{x_3}\bra{x_3} \eeq
i.e. the state under investigation is
\beq \rho = p\; \ket{\omega}\bra{\omega} + (1-p)\;\rho_{mix}\;.\label{mixing} \eeq
\\

\subsection{GHZ-like state ($\Delta = 0$)\label{sec_ghz}}
\begin{figure}[ht!]\centering\includegraphics[width=8cm]{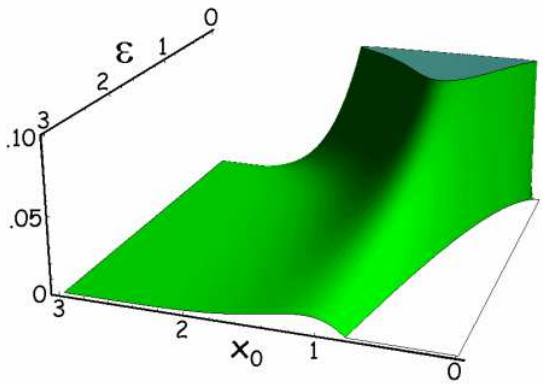}\caption{(Colour online) Illustration of the detection parameter space of the criterion (\ref{ineq}) for $k=2$ and the state $\ket{\omega}$, Eq.~(\ref{ghzstate}), with $p = 1$ and $\Delta = 0$. The state is detected to be genuinely tripartite entangled wherever the graph is above zero (i.e. in the green plotted region). It can be seen that the optimal choice for $x_0$ is between 0.7 and 1.2, depending only slightly on the value of $\epsilon$.}\label{fig_x0}\end{figure}

\begin{figure}[ht!]\centering\includegraphics[width=3.8cm]{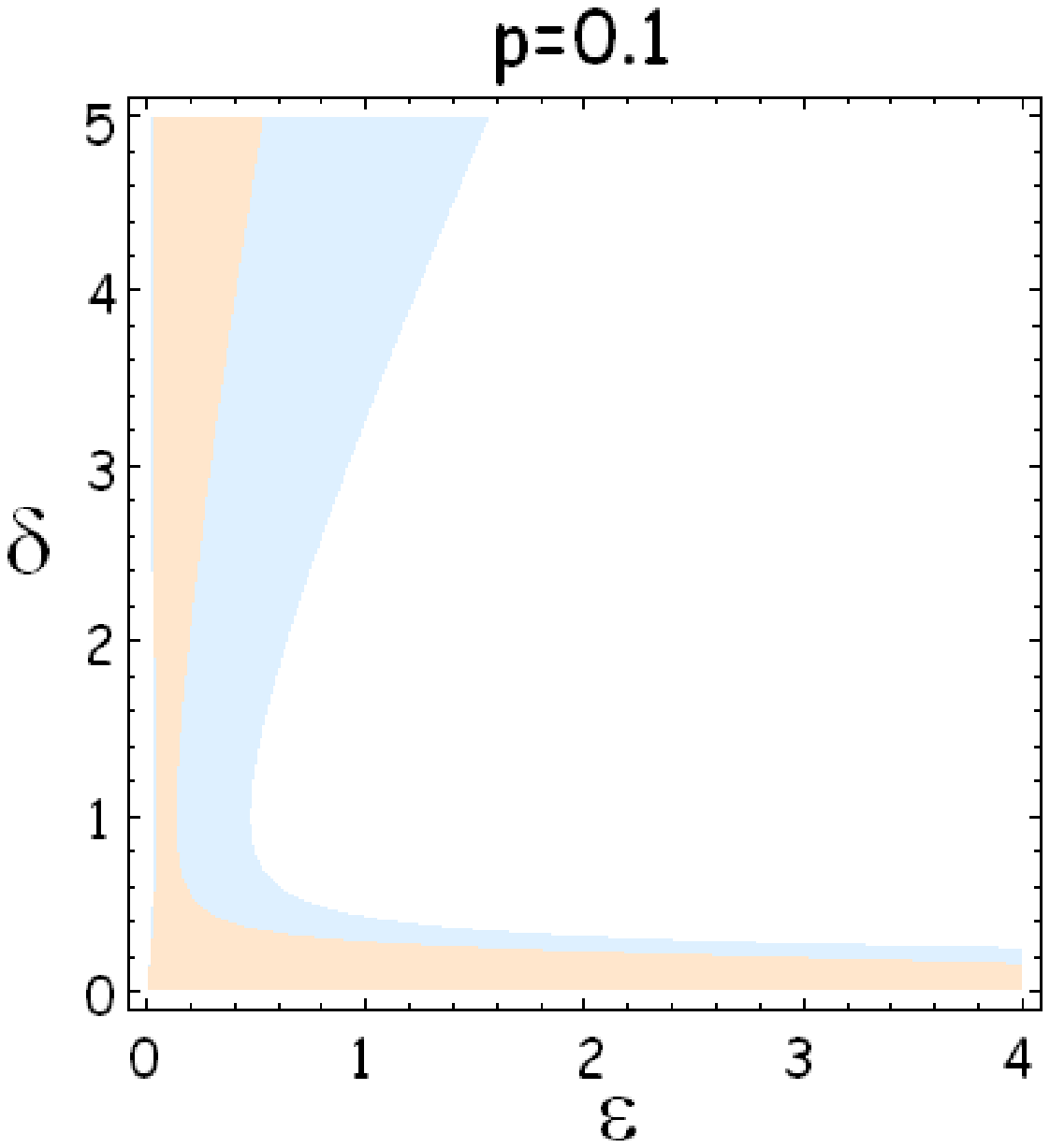}\includegraphics[width=3.8cm]{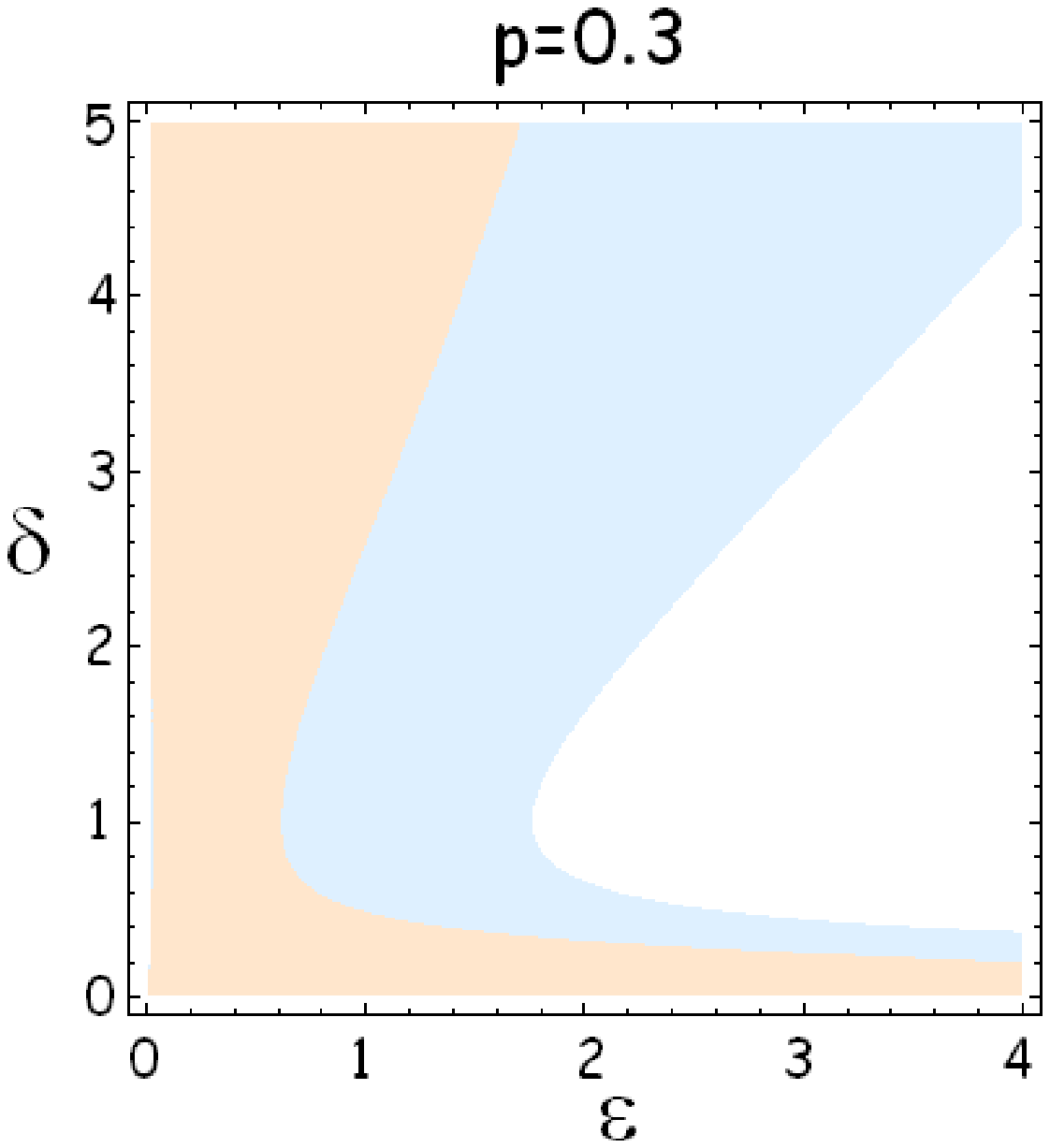}\includegraphics[width=3.8cm]{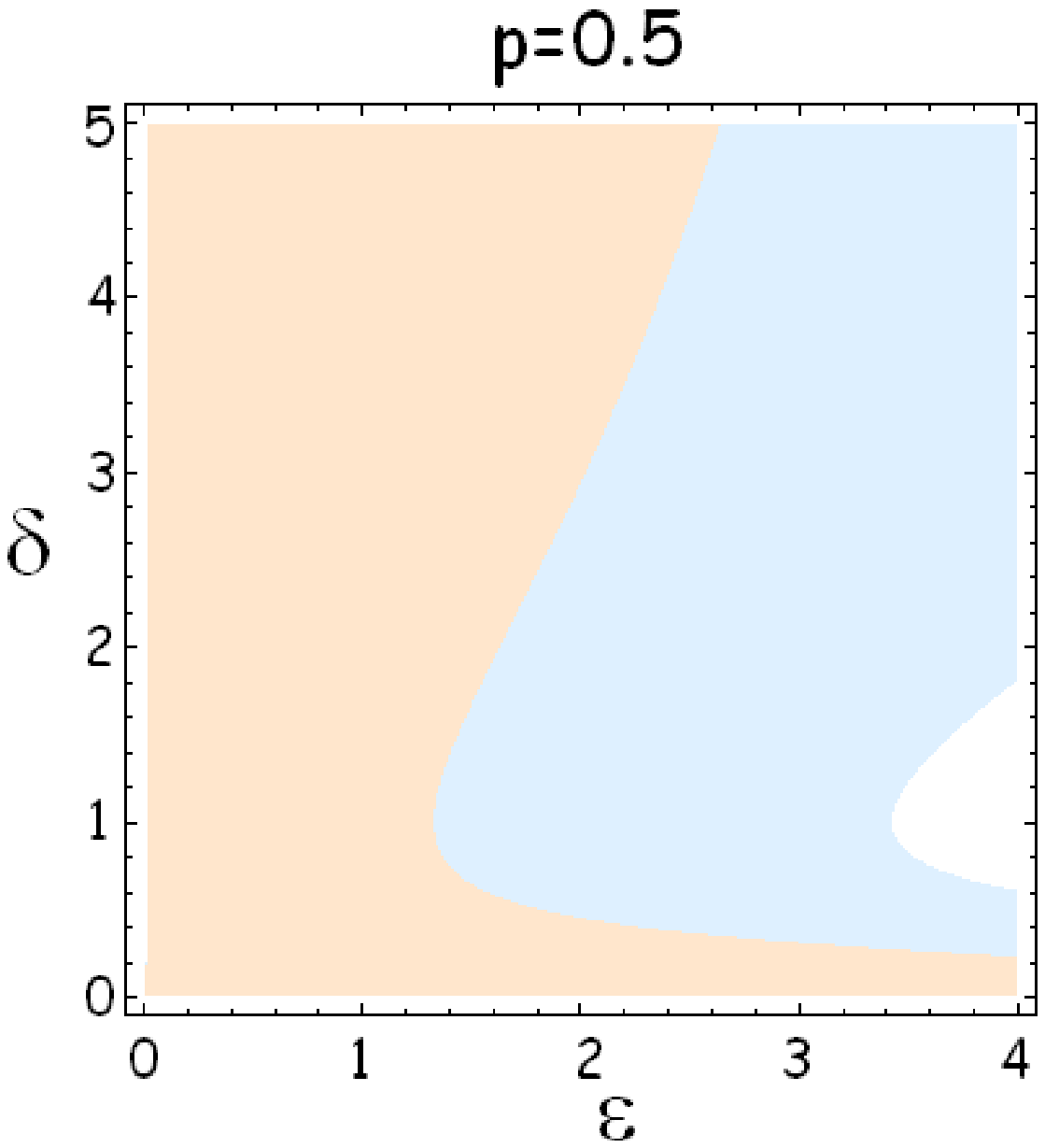}\includegraphics[width=3.8cm]{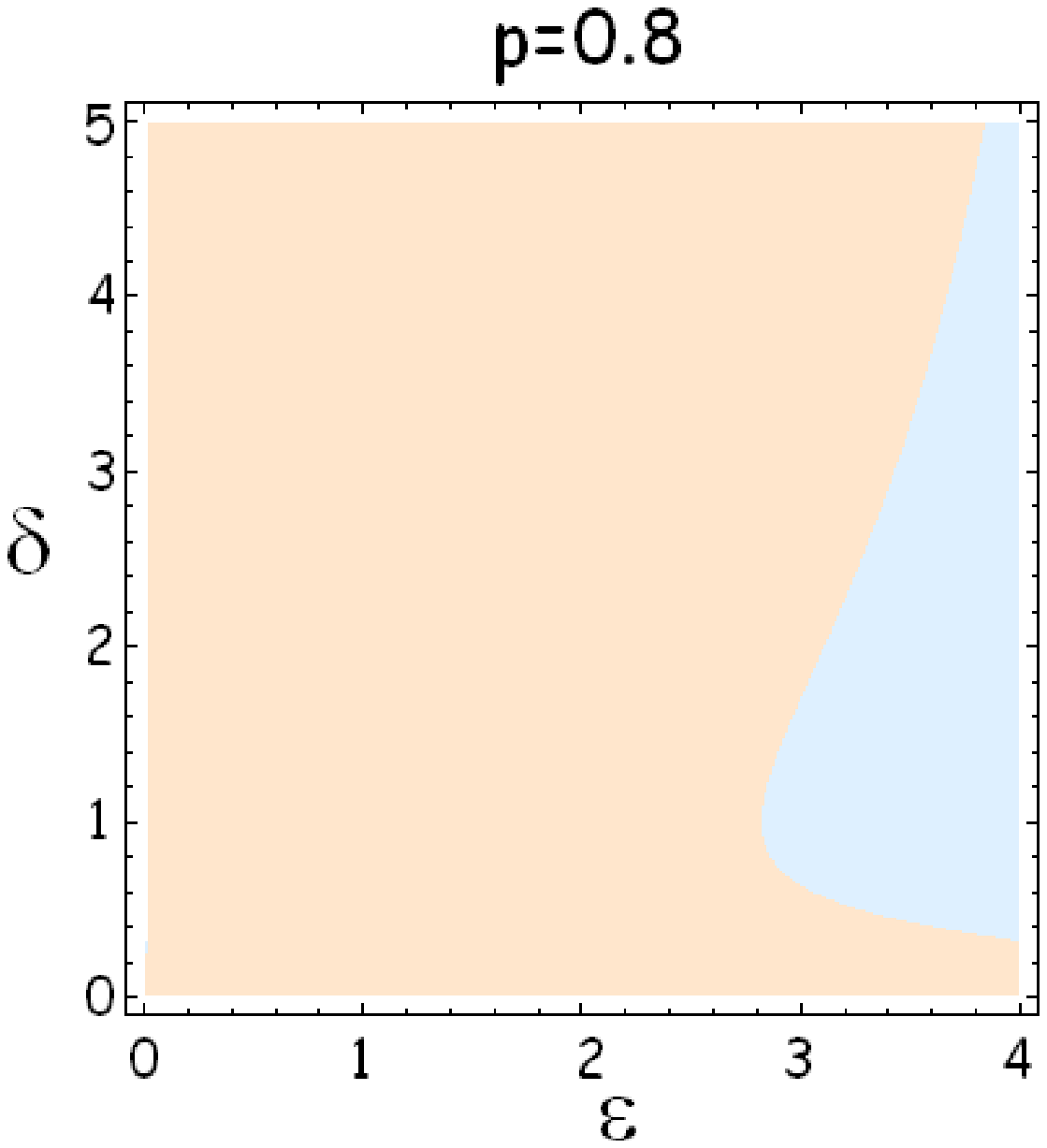}\caption{(Colour online) Illustration of the detection quality of ineq. (\ref{ineq}) for $k=2$ (red) and $k=3$ (blue) for the state (\ref{ghzstate}) for $\sigma=1, x_0=1$ and different values of $p, \epsilon$ and $\delta$. For $p = 1$ the whole state space is detected to be entangled ($k\leq 2$) and to be genuinely multipartite entangled ($k=1$) for $\epsilon < 4.648$. For lower values of $p$, still very large portions of the state space are detected to be genuinely multipartite entangled (red regions, $k=2$) or at least entangled (red and blue regions $(k\geq2)$).}\label{fig_ghzlike}\end{figure}

In the case $\Delta = 0$, the state $\ket{\omega}$ assumes the comparatively simple form
\beq \ket{\omega} = \frac{1}{\sqrt{N}} \int_{-\infty}^{\infty} d^3x \ e^{-\frac{x_1^2}{2 \sigma}} e^{-\frac{(x_1-x_2)^2+(x_1-x_3)^2}{2 \epsilon}} \ket{x_1}\otimes\ket{x_2}\otimes\ket{x_3} \label{ghzstate} \eeq
As this state's distribution function has its peak at $x_1=x_2=x_3=0$ and is correlated such that all three coordinates are most likely to be very close to each other (or even equal for $\epsilon \rightarrow 0$), $\ket{\Phi}$ is best chosen to be of the form
\beq \label{phighz} \ket{\Phi} &=&\ket{\phi_1}\otimes\ket{\phi_2}\qquad\textrm{with}\nonumber\\
 &&\ket{\phi_1}=\ket{x_0}\otimes\ket{x_0}\otimes\ket{x_0} \\ \nonumber
 &&\ket{\phi_2}=\ket{-x_0}\otimes\ket{-x_0}\otimes\ket{-x_0}  \eeq
for some $x_0 \neq 0$ (due to condition ${\cal C}2$). The optimal value for $x_0$ can be obtained from analytical or numerical optimisation. Since for $\sigma \rightarrow 0$ this state becomes separable (as can be seen directly by substituting eq. (\ref{deltafkt}) in eq. (\ref{ghzstate})), we are only interested in $\sigma > 0$ and thus can, without loss of generality, define $\sigma = 1$ and measure all lengths in units of $\sqrt{\sigma}$.\\
In order to estimate the best value for $x_0$, we first assume a pure state, i.e. $p=1$ and investigate the detection behaviour of the criterion (\ref{ineq}) for different values of the two remaining parameters $\epsilon$ and $x_0$ (as illustrated in Fig. \ref{fig_x0}). It can be seen that the optimal choice of $x_0$ depends on $\epsilon$ only slightly and is best chosen in the range $0.7 \leq x_0 \leq 1.2$. In the further study of this state, we will chose $x_0 = 1$, such that only parameters of the state remain.\\
Now, we can investigate the mixed state case (i.e. $p < 1$). In Fig.~\ref{fig_ghzlike} the detection range of our criterion (\ref{ineq}) for $k=2$ and $k=3$ is illustrated
.\\

\subsection{W-like state ($\Delta = 1$)}
\begin{figure}[ht!]\centering\includegraphics[width=3.8cm]{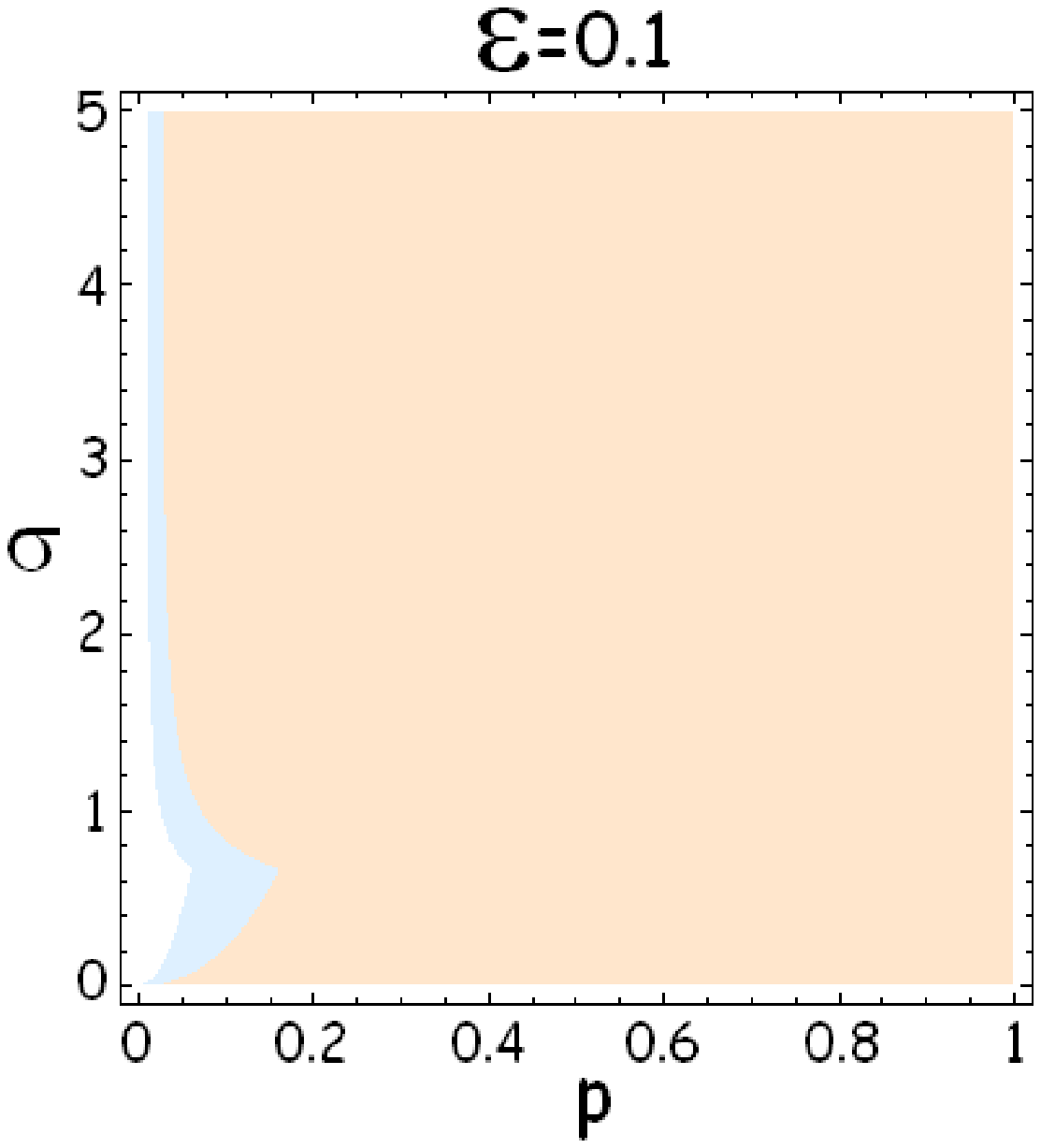}\includegraphics[width=3.8cm]{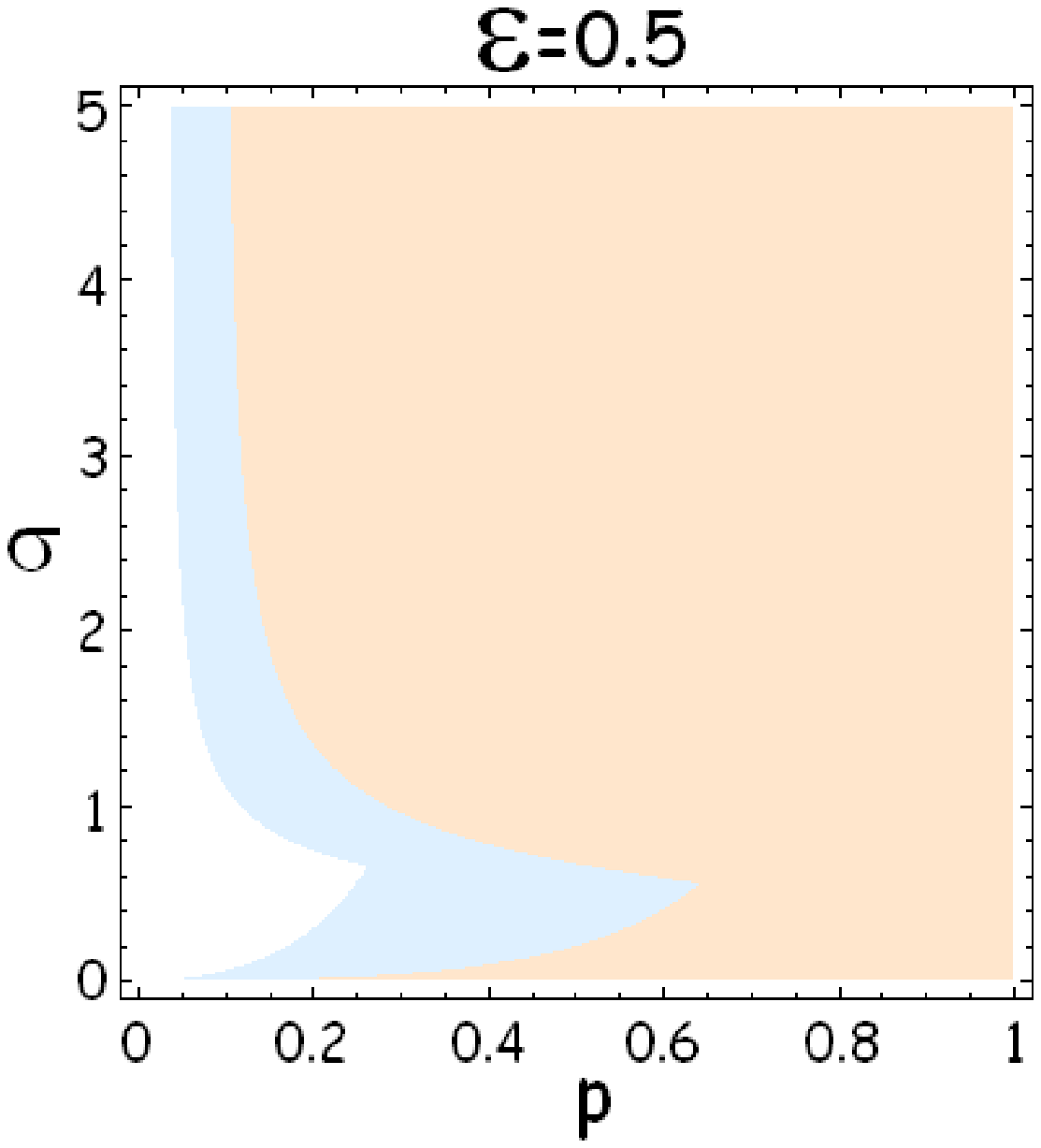}\includegraphics[width=3.8cm]{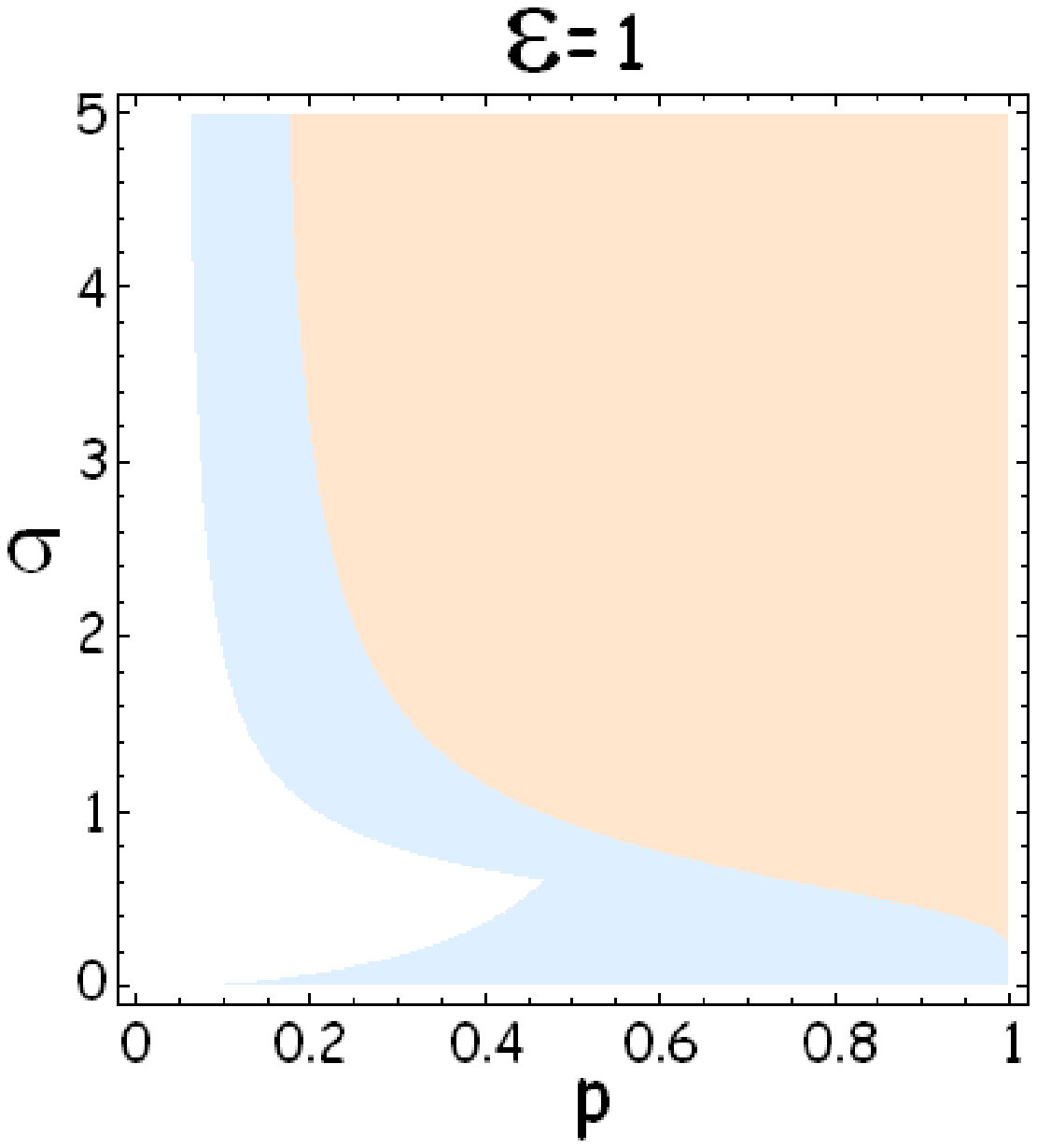}\includegraphics[width=3.8cm]{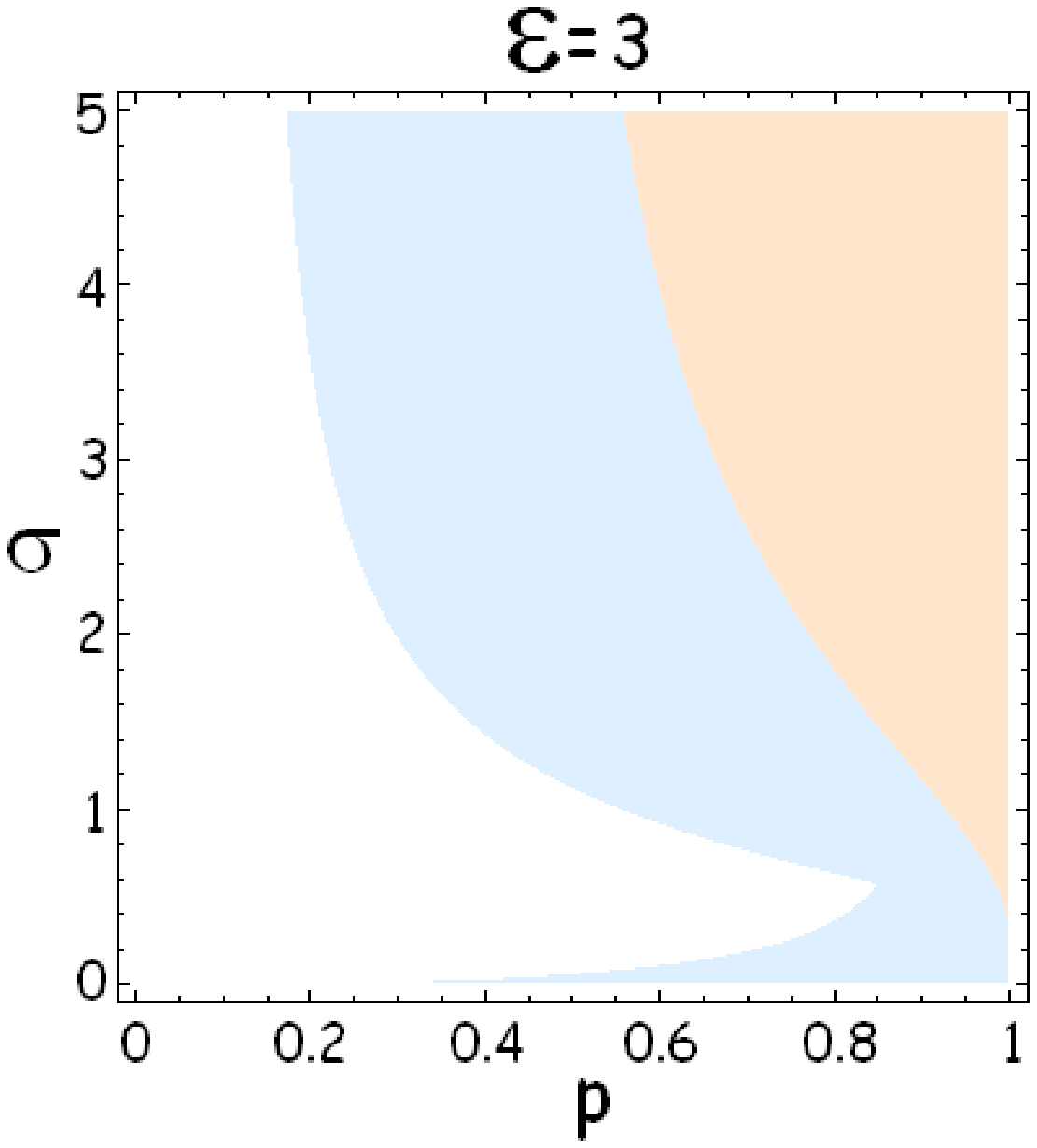}\caption{(Colour online) Detection quality of the inequality (\ref{ineq}) for the family of the tripartite states  (\ref{statedef}) with $\Delta=1, \delta=1$ and $x_0=\{0, 1.5\}$ (where $x_0 = 0$ detects better if $\sigma < 1$, even for very small $p$, but ceases to detect for large $\epsilon$; and $x_0 = 1.5$ detects better for $\sigma > 1$ and still detects a considerable amount of entanglement for large $\epsilon$). The red area is detected to be genuinely multipartite entangled (2-inseparable) and the blue and red areas are detected to be entangled (3-inseparable).}\label{fig_wlike}\end{figure}

Let us now investigate a more general state, namely the case $\Delta \neq 0$. Without loss of generality, we can set $\Delta = 1$ and thus measure all lengths in units of $\Delta$. The optimal choice of $\ket{\Phi}$ of course has to be a generalisation of the one used in the previous section, which coincides with it for $\Delta = 0$. The choice is quite straightforward, since the state (\ref{statedef}) contains six terms. Combination of any two of them leads to the desired high-magnitude off-diagonal density matrix element, the only restrictions being conditions ${\cal C}1$-${\cal C}3$. This leads to
\beq \ket{\Phi} &=&\ket{\phi_1}\otimes\ket{\phi_2}\qquad\textrm{with}\nonumber\\
 &&\ket{\phi_1}= \ket{x_0+\Delta}\otimes\ket{x_0}\otimes\ket{x_0-\Delta}\\
 &&\ket{\phi_2}=\ket{-x_0-\Delta}\otimes\ket{-x_0+\Delta}\otimes\ket{-x_0} \nonumber \eeq
In this rather complicated case, numerical optimisation is necessary for achieving optimal detection quality, since the optimal $x_0$ depends strongly on the other parameters (more than in the previous case with $\Delta=0$). However, even without numerical optimisation and using only two different choices of $x_0$ (namely $x_0 = 0$ and $x_0 = 1.5$) the detection range of our criterion (\ref{ineq}) is very wide, as illustrated in Fig.~\ref{fig_wlike}.\\

\subsection{Non-Gaussian States}
\begin{figure}[ht!]\centering\includegraphics[width=8cm]{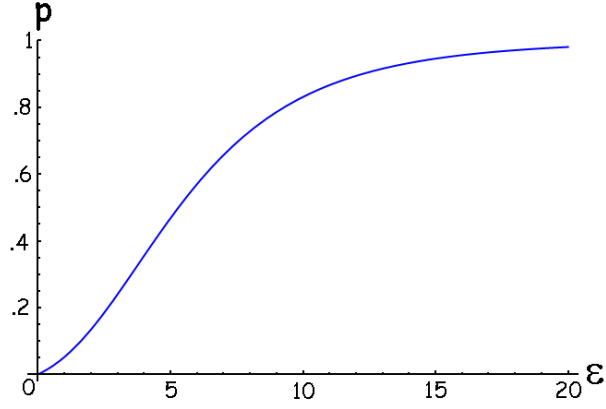}\caption{(Colour online) Illustration of the detection parameter space of the criterion (\ref{ineq}) for $k=2$ and the state $\rho$, Eq.~(\ref{mixing}) where $\ket{\omega}$ as in (\ref{nongaussexpstate}), with $\sigma = 1$ and $\delta = 3$. The curve indicates the critical proportion $p$, i.e. the detection threshold for genuine multipartite entanglement.}\label{fig_nongauss}\end{figure}

The above two examples belong to the class of Gaussian states, which have always been the main focus of research in the field of continuous variable entanglement (see e.g. Refs.~\cite{giedke, adesso06, adesso08, vloock}). However, recently questions regarding non-Gaussian entangled states have attracted a lot of interest within the scientific community. Since the criterion presented in this paper is - unlike most other criteria for continuous variable entanglement - not tailored specifically for Gaussian states, we will also show its detection quality for non-Gaussian states.\\
Consider the state
\beq \rho = p \ket{\omega}\bra{\omega} + (1-p)\rho_{mix} \eeq
where
\beq
\ket{\omega} = \frac{1}{N_1} \int d^3x \Theta(\beta-|x_1|) \Theta(\epsilon-|x_1-x_2|) \Theta(\epsilon-|x_1-x_3|) \ket{x_1}\ket{x_2}\ket{x_3} \\
\rho_{mix} = \frac{1}{N_2} \int d^3x \Theta(\delta-|x_1|) \Theta(\delta-|x_2|) \Theta(\delta-|x_3|) \ket{x_1}\bra{x_1} \otimes \ket{x_2}\bra{x_2} \otimes \ket{x_3}\bra{x_3} \eeq
with
\beq N_1 = 8 \epsilon^2\beta \quad N_2 = 8 \delta^3 \quad \beta,\epsilon,\delta > 0 \eeq
where $\Theta(x)$ is the Heaviside-function. This state is a non-Gaussian modification of the GHZ-like state (\ref{ghzstate}).\\
Using (\ref{phighz}) and chosing $x_0$ appropriately, the criterion (\ref{ineq}) reads
\beq 0 \leq \left\{ \begin{array}{ccc} 
\frac{p}{8\epsilon^2\beta} \
\quad \quad \quad \mathrm{if} \ \frac{\epsilon}{2} < \beta \ \mathrm{and} \ \delta < \beta \\
\frac{p}{8\epsilon^2\beta} - \frac{1-p}{8\delta^3} \quad \quad \mathrm{if} \ \frac{\epsilon}{2} < \beta \leq \delta \\
0 \quad \quad \quad \quad \mathrm{else} \end{array} \right. \eeq
where the first condition is always positive for $p>0$ and thus indicates genuine multipartite entanglement, the yield of the second condition depends on the parameters and the third condition can never be violated. In particular, the first condition detects a large portion of the state space to be genuinely multipartite entangled already for infinitesimal $p>0$.\\
\\
Since the state discussed above is only a rather simple example and not very close to experimental realisation, let us illustate the detection quality of our criterion for another non-Gaussian state, which is more likely to be implemented experimentally \cite{nongauss1, nongauss2}, namely
\beq \ket{\omega} = a_1a_2a_3 \frac{1}{\sqrt{N}} \int_{-\infty}^{\infty} d^3x \ e^{-\frac{x_1^2}{2 \sigma}} e^{-\frac{(x_1-x_2)^2+(x_1-x_3)^2}{2 \epsilon}} \ket{x_1}\otimes\ket{x_2}\otimes\ket{x_3} \label{nongaussexpstate}\eeq
where the $a_i$ are annihilation operators and
\beq N = \frac{1}{8} \pi^{3/2} \epsilon \sigma^{3/2} (\epsilon^2+6\epsilon\sigma +15\sigma^2) \eeq
mixed with Gaussian noise, as in (\ref{mixing}) with (\ref{rhomix}). This state also represents a modification of (\ref{ghzstate}) and can therefore be detected by the same choice of $\ket{\Phi}$.\\
Without loss of generality, we set $\sigma = 1$ and thus measure all lengths in units of $\sqrt{\sigma}$. For $\delta \leq \frac{3\sigma}{2}$, the whole state space with $p>0$ is detected to be genuinely multipartite entangled, independantly of $\epsilon$. For $\delta > \frac{3\sigma}{2}$, still large areas of the state space are detected, as illustrated in Fig. \ref{fig_nongauss}.

\section{Experimental Implementation\label{sec_exp}}
Since it is quite important for multipartite entanglement criteria to not only work in theory but also to be implementable experimentally (without resorting to a full state tomography, since this would mean an infinite number of required measurements), let us illustrate how this can be done for our criterion, ineq. (\ref{ineq}), for the pure example state from section \ref{sec_ghz}, the $GHZ$--like state (\ref{ghzstate}). Generalisation to other states is straightforward, but might be rather cumbersome.\\
While the criterion detects optimally for sharp $\ket{\Phi}$, such states cannot be measured physically, since detectors always have a finite size. We will thus assume
\beq \ket{\Phi}=\ket{\alpha}\ket{\alpha}\ket{\alpha}\ket{\beta}\ket{\beta}\ket{\beta} \eeq
with
\beq \ket{\alpha} = \frac{1}{\sqrt{\xi}} \int dx \ \Theta\left(x-x_0+\frac{\xi}{2}\right) \Theta\left(x_0+\frac{\xi}{2}-x\right) \ket{x} \\ \nonumber \ket{\beta} = \frac{1}{\sqrt{\xi}} \int dx \ \Theta\left(x+x_0+\frac{\xi}{2}\right) \Theta\left(-x_0+\frac{\xi}{2}-x\right) \ket{x} \eeq
where $\Theta(x)$ is the Heaviside distribution and $\xi$ is the size of the detector, for example a charge-coupled device (CCD). The $(2^n-1)$ scalar products needed for computation of ineq. (\ref{ineq}) thus take forms like
\beq \nonumber \bra{\phi_1}\rho\ket{\phi_2} = \left(\frac{\pi \epsilon}{2 \xi^{\frac{3}{2}}}\right)^2 && \int_{x_0-\frac{\xi}{2}}^{x_0+\frac{\xi}{2}}dx \ e^{-\frac{x^2}{2\sigma}} \left(\textrm{Erf}\left(\frac{2x_0-2x+\xi}{2 \sqrt{2\epsilon}}\right) + \textrm{Erf}\left(\frac{2x_0-2x-\xi}{2 \sqrt{2\epsilon}}\right)\right)^2 \cdot \\ &&
\int_{x_0-\frac{\xi}{2}}^{x_0+\frac{\xi}{2}}dy \ e^{-\frac{y^2}{2\sigma}} \left(\textrm{Erf}\left(\frac{2x_0-2y+\xi}{2 \sqrt{2\epsilon}}\right) + \textrm{Erf}\left(\frac{2x_0-2y-\xi}{2 \sqrt{2\epsilon}}\right)\right)^2 \\ \nonumber
\bra{\Phi}P_1^\dagger\rho P_1\ket{\Phi} = \left(\frac{\pi \epsilon}{2 \xi^{\frac{3}{2}}}\right)^2 && \left(\int_{-x_0-\frac{\xi}{2}}^{-x_0+\frac{\xi}{2}}dx \ e^{-\frac{x^2}{2\sigma}} \left(\textrm{Erf}\left(\frac{2x_0-2x+\xi}{2 \sqrt{2\epsilon}}\right) + \textrm{Erf}\left(\frac{2x_0-2x-\xi}{2 \sqrt{2\epsilon}}\right)\right)^2\right)^2
\eeq
where
\beq \textrm{Erf}(a) = \frac{2}{\sqrt{\pi}} \int_0^a dx \ e^{-x^2} \eeq
is the Gaussian error function. These integrals can easily be computed numerically, given the parameters used in the experimental setup, which allows for a simple prediction of measurement outcomes. We now explicitly show how to write ineq. (\ref{ineq}) in terms of expectation values of local observables in the exemplary three particle case. To that end let us first define the following local observables constructed from the finite sized detectors
\begin{eqnarray}
\sigma_x=|\alpha\rangle\langle\beta|+|\beta\rangle\langle\alpha|\nonumber\\
\sigma_y=i|\alpha\rangle\langle\beta|-i|\beta\rangle\langle\alpha|\\
\sigma_z=|\alpha\rangle\langle\alpha|-|\beta\rangle\langle\beta|\nonumber\,,
\end{eqnarray}
which are the Pauli operators of the two dimensional subspace spanned by $|\alpha\rangle$ and $|\beta\rangle$. Using the short hand notation
\begin{equation}
ijk:=\langle\sigma_i\otimes\sigma_j\otimes\sigma_j\rangle_\rho\, ,
\end{equation}
where $\sigma_1:=\mathbbm{1}$, we can explicitly rewrite ineq.(\ref{ineq}) for $k=2$ and $n=3$ as
\begin{eqnarray}
&&\frac{1}{8}|xxx-yyx-yxy-xyy+i(yyy-xxy-xyx-yxx)|-\nonumber\\
&&\frac{1}{8}(\sqrt{(111+zz1-z1z-1zz+11z-1z1-z11+zzz)(111+zz1-z1z-1zz-11z+1z1+z11-zzz)}+\\
&&\sqrt{(111-zz1+z1z-1zz+11z-1z1+z11-zzz)(111-zz1+z1z-1zz-11z+1z1-z11+zzz)}+\nonumber\\
&&\sqrt{(111-zz1-z1z+1zz+11z+1z1-z11-zzz)(111-zz1-z1z+1zz-11z-1z1+z11+zzz)})\nonumber\leq 0\,.
\end{eqnarray}
It is also possible to decompose the inequalities in terms of local expectation values for larger $n$ or different $k$ in a straightforward way. This however yields rather cumbersome expressions, which is why they are not presented then here in full detail.\\
\\
Experimental measurement uncertainties can be estimated by means of the Gaussian law of error propagation, which states that the measurement uncertainty $\Xi$ of a function $f$ of several arguments $x_i$ is given by
\beq \Xi = \sqrt{\sum_i \left(\frac{\partial f}{\partial x_i} \zeta_i \right)} \eeq
where $\zeta_i$ are the respective measurement uncertainties of the $x_i$.\\
We will assume that all expectation values $x_i = \bra{\alpha_i}\rho\ket{\alpha_i}$ underlie the same relative uncertainty (i.e. $\zeta_i / \bra{\alpha_i}\rho\ket{\alpha_i} = \zeta$ is independant of $i$), such that only two uncertainty parameters remain, namely $\zeta$ and $o$, the latter being the absolute uncertainty of the first term in ineq. (\ref{ineq}). Now, the measurement uncertainty $\Xi$ of the whole inequality is given by
\beq \Xi^2 = o^2 + \sum_{\alpha,i} \left( \frac{1}{2k} \prod_{j=1}^{2k} \frac{(x_j)^(1/2k)}{x_i} \zeta_i \right)^2 = o^2 + \frac{1}{4k^2} \sum_{\alpha,i} \prod_{j=1}^{2k} (x_j)^{1/k} \zeta^2 \leq o^2 + \frac{\zeta^2 \gamma}{8k^3} \label{error} \eeq
where the ineqality follows from the fct that a geometric mean is maximal whenever all its factors are equal, and
\beq \gamma = \sum_{l=1}^k \frac{(-1)^{k-i} i^{n-1}}{(i-1)! (k-i)!} \eeq
is the number of all $k$-partitions of an $n$-partite system.\\
In our above examples (i.e. if $n=3$ and $k = 2$ or $k=3$), the second term in (\ref{error}) is much smaller than the first, such that the measurement uncertainty of the complete expression (\ref{ineq}) is approximately equal to the uncertainty of its first term: $\Xi \approx o$, which makes this kind of experimental uncertainty easy to deal with.\\
\\
Another kind of complication that is to be expected in experimental realisations (e.g. in quantum optics) are imperfect detectors. These correspond to nonperfect projective measurements, i.e. each scalar product $\bra{\alpha}\rho\ket{\beta}$ is multiplied by some factor $0 \leq \tau \leq 1$. Since the whole inequality is now linear in $\tau$, this does not alter the violation or nonviolation of the inequality, but reduces the magnitude of violation linearly.\\

\section{Summary}
We present a criterion for $k$-separability in multipartite continuous variable systems. It is an inequality which is satisfied for all $k$--separable states, i.e. any violation implies that the state is not $k$--separable. The criterion particularly allows to distinguish between biseparable states (which can only be used in few applications) and genuinely multipartite entangled ones (which are a basic building piece for several applications of quantum information theory which go beyond the potential of classical systems). We show how the inequality can be optimised by chosing an appropriate state $\ket{\Phi}$ (for which we give four explicit conditions) and thus being left with a reduced (finite) number of optimisation parameters, which can be computed. We analyse two different families of states, which may be considered to be generalisations of the most famous genuinely multipartite states in finite quantum information theory, the $GHZ$-type entangled states and the $W$-type entangled states. Our criterion easily detects a large parameter space of entangled states when mixed with Gaussian noise. Since no comparable criteria exist, it can not be said how tight these detection thresholds are.\\
Moreover, we explicitly show how the developed criterion for $k$-separability in multipartite continuous variable systems can be rewritten by local expectations values, thus how this criterion can be experimentally implemented.\\
In summary, we presented a computable criterion for detecting $k$-inseparability (and particularly genuine multipartite entanglement) in continuous variable systems which can be experimentally implemented by finitely many local observables.\\

\section{Acknowledgements}
The authors would like to thank Gerardo Adesso and Paul Erker for productive and useful discussions. Andreas Gabriel and Marcus Huber gratefully acknowledge the Austrian Fund project FWF-P21947N16. Andreas Gabriel is supported by the University of Vienna’s research grant. Beatrix C. Hiesmayr acknowledges the EU project QESSENCE.

\end{document}